\documentclass[11pt,a4paper]{amsart}
\usepackage[utf8]{inputenc}
\usepackage[normalem]{ulem}
\usepackage{bbm}
\usepackage{lmodern,bm}
\renewcommand{\mathbf}{\bm}
\usepackage{color}
\usepackage{graphicx}
\usepackage{graphics}

\numberwithin{equation}{section} 

\newtheorem{example}{Example}
\newtheorem{remark}{Remark}

\newcommand{\hA}{\mathcal{A}}
\newcommand{\hB}{\mathcal{B}}

\newcommand{\hE}{\mathcal{E}}

\newcommand{\hH}{\mathcal{H}}
\newcommand{\hI}{\mathcal{I}} 

\newcommand{\hK}{\mathcal{K}}
\newcommand{\hL}{\mathcal{L}}
\newcommand{\hM}{\mathcal{M}}

\newcommand{\hP}{\mathcal{P}} 

\newcommand{\hR}{\mathcal{R}}

\newcommand{\sfe}{\mathsf{E}}
\newcommand{\sff}{\mathsf{F}}

\newcommand{\sfm}{\mathsf{M}}
\newcommand{\sfp}{\mathsf{P}}
\newcommand{\sfq}{\mathsf{Q}}
\newcommand{\sfz}{\mathsf{Z}}

\newcommand{\R}{\mathbb R} 
\newcommand{\C}{\mathbb C} 
\newcommand{\Q}{\mathbb Q} 

\newcommand{\fii}{\varphi} 

\newcommand{\lh}{\mathcal{L(H)}} 

\newcommand{\tr}[1]{\mathrm{tr}\left[#1\right]} 
\def\<{\langle} 
\def\>{\rangle} 
\newcommand{\kb}[2]{|#1 \rangle\langle #2|} 
\newcommand{\ip}[2]{\left\langle #1 | #2 \right\rangle} 

\newcommand{\br}{\mathcal B(\mathbb R)} 

\newcommand{\ba}{\begin{array}}
\newcommand{\ea}{\end{array}}

\begin{document}
\title[Relational interpretation]{An attempt to understand\\ relational quantum mechanics} 

\author{Pekka Lahti}
\address{University of Turku, Turku, Finland}
\email{pekka.lahti@utu.fi}
\author{Juha-Pekka Pellonp\"a\"a}
\address{Department of Physics and Astronomy, University of Turku, Turku, Finland}
\email{juhpello@utu.fi}
\begin{abstract} We search for a possible mathematical formulation
of some of the key ideas of the relational interpretation of quantum mechanics and study their consequences. We also briefly overview some proposals of relational quantum mechanics for an  axiomatic reconstruction of the Hilbert space formulation of quantum mechanics.
\end{abstract}

\maketitle

\thispagestyle{empty}



\section{Introduction}
This is an attempt to understand Rovelli's relational interpretation of quantum mechanics, RQM, as  outlined in \cite{Rovelli1996,Smerlak_Rovelli2007, Laudisa_Rovelli2019, Rovelli2018, Martin_Rovelli_etal_2019}, further advocated in \cite{Rovelli2021} and sharpened in \cite{Adlam_Rovelli2022}\footnote{We are grateful to Carlo Rovelli for informing us on the recent  paper \cite{Adlam_Rovelli2022}  which contains a slightly modified  list of the defining assumptions of RQM.}.
We try to identify and express some of  the key ideas of RQM  within  a systematic Hilbert space formulation of quantum mechanics.
RQM follows the old idea 
that properties of things are relational, and they get actualized in interactions. This poses the question how to justify this point of view  within the Hilbert space structures of  quantum mechanics 
without facing the well-known measurement problem.

 Our study is in line   with the critical analyses of \cite{BvanF2010,Ruyant2018,Laudisa2019,Pienaar2021,Brukner2021} though our empahasis is more on the formal side of scrutinizing   the assumptions constituting RQM, as we read them.\footnote{Though the RQM literature is already quite abundant, it seems to us that
no common understanding of this interpretation has  yet been reached. This may be due to the lack of  formally rigorous presentation of the basic assumptions of the interpretation.} 
We also consider briefly the possible role of the basic ideas of RQM  in an
 axiomatic reconstruction of (the Hilbert space formulation of) quantum mechanics.

Several authors have  examined  the  relational interpretation of quantum mechanics.
 Especially, as Rovelli writes in his excellent little book {\em Helgoland: making sense of the quantum revolution} \cite[p.\ 142]{Rovelli2021}, "the world of philosophy has reacted to this interpretation in various ways: different schools of thought have framed it in different philosophical terms."  \,  From  \cite[p.\ 8]{Rovelli2018} we may also read that "[t]here are several objections that come naturally to mind when one first encounters relational QM, which seem to render it inconsistent.
These have been long discussed and have all been
convincingly answered; \ldots 
I will not rediscuss them here. Relational QM is a consistent
interpretation of quantum theory."\,\footnote{In a recent paper Di Biagio and Rovelli   \cite{Rovelli_DiBiagio2021} also write:
"In recent works, \v Caslav Brukner and Jacques Pienaar have raised interesting objections to the  relational interpretation of quantum mechanics. We answer these objections in detail and show that, far from questioning the viability of the interpretation, they sharpen and clarify it."} Still, we dare to take our go with it, with the hope that our analysis could contribute to a further clarification of the content of  RQM.

We try to  identify step by step, in the order of increasing requisite, the assumptions which seem to constitute  the relational interpretation.  
The key notions, as we see them, are: property (a value of a physical quantity) (sec.\ \ref{GB}),
interaction  (sec.\ \ref{interaction}), strong correlation (sec.\ \ref{correlation}), perspective as a measurement scheme (sec.\ \ref{scheme}), repeatability (sec.\ \ref{repeatability}), and local collapse  (sec.\ \ref{RQMlc}). 

Physical quantities and their possible values, interactions, correlations, measurement schemes and repeatable measurement schemes are standard ingredients of quantum mechanics, whereas local collapse (our term) is a specialized form of the controversal projection postulate.
In a nutshell, in our analysis RQM consists of the statement that physical quantities take values  through their repeatable measurement schemes together with an {\em ad hoc} local collapse to circumvent the objectification problem.

In applying the above set of rules  in the context of  sequential and joint measurement schemes, sec.\ \ref{RQMsp}, we come at  odds with a claim of  RQM that "we all `see the same world'"\cite[sec.\ 2.6]{Laudisa_Rovelli2019}.
In \cite{Adlam_Rovelli2022}, however,   
relying on the analysis of Di Biagio and Rovelli  \cite{Rovelli_DiBiagio2021},   Adlam and Rovelli
supplement RQM with a new postulate called {\em cross-perspective links} to reach intersubjectivity between `all parties'. 
Though remedying the possible  lack  of objectivity of the description,  our reading of it 
in sec.\ \ref{CPL} rather suggests
that this postulate is  a disguised form of the projection postulate leading to a `global collapse' of the state of the interacting pair.

In the proposed  axiomatic reconstruction of RQM, two postulates have been formulated to outline  the basic event structure of the theory. Though interesting, these assumptions constitute only a beginning of an axiomatic reconstruction of the Hilbert space formulation of quantum mechanics. It remains to be seen if  such a reconstruction could be completed to yield a (version of the)  relational interpretation of quantum mechanics.

\section{Hilbert space theory and RQM} 
Relational quantum mechanics does not intend to change the mathematical apparatus  of quantum mechanics, but proposes, in particular,  a way of attaching values to the  variables of a physical system relative to other systems interacting with it. 
RQM also treats all physical systems, including measuring apparata and  even observers, equally on the same basis as quantum systems.  
Therefore, it is safe to start with  the Hilbert space formulation of quantum mechanics, together with a full use of its theory of compound systems within which mutually interacting systems are described.\footnote{In \cite{Adlam_Rovelli2022} it is explicitly stated that "[RQM]  does not require us to add anything to the existing mathematical framework of quantum mechanics."  \, The second postulate of RQM even states that "unitary quantum mechanics is complete", {\em ibid.}}

\subsection{General background}\label{GB}
To fix our notations, consider a physical system $S$ with its Hilbert space $\hH$, complex  and separable.\footnote{We use {\em Quantum Measurement}  \cite{QM} as a basic reference to the standard  results  used in this writing.}
Its  physical quantities (variables,  observables, or whatever term one prefers to use)  are given (represented) as the  normalized positive operator measures (semispectral measures) $\sfe$ defined on $\sigma$-algebras $\hA$ of subsets of  sets $\Omega$ and taking values in the set $\hE(\hH)\subset\hL(\hH)$ of effects, i.e.\ positive operators bounded above by the identity $I$; here $\hL(\hH)$ is the set of the bounded operators on $\hH$. Intuitively, such sets $\Omega$ consist of the possible values or measurement outcomes of the observables whereas the sets $X\in\hA$ are the bin sets within which the values or outcomes are noticed, observed, or registered. Typically, $(\Omega,\hA)$ is the real Borel space $(\R,\br)$, or an appropriate subspace of it.

Among the observables $\sfe:\hA\to\lh$ there are the (real)  spectral measures $\sfp:\br\to\lh$, taking projection operators as values, and which are uniquely associated with the selfadjoint operators $A=\int x\,d\sfp(x)$. Occasionally, we may refer to such observables as the sharp observables. Apart from their central importance, there is, however,  no point to restrict attention  to such observables only, and still less so to the observables given as discrete selfadjoint operators (with the spectral structure $A=\sum a_i P_i$).\footnote{Such a restriction would only hide some of the crucial assumptions concerning the possible measurements or  value assignments of such observables.} 

The other fundamental notion of the theory is the notion of  a state as  a positive trace class operator of trace one,  $\rho:\hH\to\hH,$ $\tr{\rho}=1$.
Any observable-state pair $(\sfe,\rho)$ defines a probability measure $p^\sfe_\rho:\hA\to[0,1]$ through $p^\sfe_\rho(X)=\tr{\rho\sfe(X)}$, with the obvious, but important properties that  if $p^\sfe_\rho=p^\sff_\rho$, for all $\rho$, then $\sfe=\sff$, and if 
$p^\sfe_\rho=p^\sfe_\sigma$, for all $\sfe$, then $\rho=\sigma$.
In particular,  states can thus be identified   with the totality of the  probabilities $p^\sfe_\rho(X)$, for all $\sfe$ and $X$.

For the pure states, the extremal elements of the convex set of states, we also use the notations
 $\rho=P[\fii]=\kb{\fii}{\fii}$, with $\fii\in\hH$ being a unit vector. Occasionally, we  may refer to pure states also as vector  states and use  unit vectors   as their representatives.

In the so-called minimal interpretation of quantum mechanics, the meaning of the number $p^\sfe_\rho(X)$ is the probability that a measurement of $\sfe$ on $S$  in state $\rho$ leads to a result in $X$, with the idea that these probabilities approximate the actual measurement outcome statistics obtained by repeating the same measurement under the same conditions many times.  As many other interpretations, RQM is not content with  this interpretation.\footnote{In the above wording, minimal interpretation does not address the question of the meaning of probability, that is, it does not take a position on the question of the interpretation of the notion of  probability.
Though a position could be taken (modal frequency,  propensity, \ldots), it is, in this context,  unnecessary  and would just  hide  our intentions. }

\begin{remark}\label{RQM1}\rm
In classical mechanics, one may consistently assume that all physical quantities (as functions on the phase space of the system) always have well-defined, though possibly unknown values (which may vary in time). Thus, if we know, in which phase space point the system is, that is, we know the values of the canonical variables, we may compute the value of any other variable. Therefore, the phase space points  may be referred to as the pure (maximal information) states of the system, other (mixed) states being expressed as probabilty measures on phase space and are  used to describe possible ignorance (of what ever source) on the actual phase space point. One may thus  assume that a classical system is always in a pure state, that is, all the variables have well-defined values,  though in the case of incomplete  information one needs to rely on  a probability distribution of the values coded in a mixed state.\footnote{The structure of the set of probability measures on phase space is that of a  (Bauer)  simplex, which allows one to express any mixed state uniquely as a `generalized convex combination' of the extreme points, the pure states. Clearly, they are the point measures identified as the phase space points.} 

In quantum mechanics one cannot consistently assume that all the quantities $\sfe$ of a system would always have well-defined values. Still, one may assume that the system is always in a well-defined state $\rho$, that is,  the system is always characterized by the totality of the probability measures $p_\rho^\sfe$. Even in case of pure states, some of those probabilities are always nontrivial ($\ne 0,1$), and in  the case of mixed states, a classical type of  ignorance interpretation of the probability measures $p^\sfe_\rho$ is ruled out by  the nonunique decomposability of mixed states into pure states.\footnote{Any mixed state has a continuum of different decompositions into pure states.  A full characterization of the possible decompositions of a mixed state is given in \cite{CDeVL1997}, see also  \cite[Thm 9.2.]{QM}.}
\qed
\end{remark}

RQM does not consider   state as a fundamental notion but still uses it exclusively in two different roles.  First, it appears as a  bookkeeping means of the probabilities $p^\sfe_\rho(X)$: 
"[state] is a theoretical device we use for bookkeeping information about the values of variables of [a physical system] $S$ actualized in interactions with [another physical system] $S'$, values which can in principle be used for predicting other (for instance future, or past) values that variables may take in other interactions with $S'$."\,  \cite[p.\ 6]{Rovelli2018}.  In 
 \cite[p.\ 4]{Rovelli2018}  it is further clarified  that technically the notion of state $\rho$ is computed from the expectation values $\tr{\rho A}$, $A$ being a (bounded) selfadjoint operator, equivalently, from the totality of the probabilities $p^\sfe_\rho(X)$, for all $\sfe$ and $X$.\footnote{Clearly, one does not need all the  observables here, but one is not enough  unless it is informationally complete (and thus necessarily given as a noncommutative semispectral measure), or   some prior information is available. 
Especially, 
the  statistics of a complete set of mutually commuting (sharp) observables does not,  in general,  suffice to determine the state of the system. However, RQM does not explain the meaning of these expectation values in that interpretation.}
The second use of this notion  is what will be called local collapse, Sec.\ \ref{RQMlc}, and it defines  the state of a system with respect to another system, after the two systems have interacted with a consequence that a physical quantity of the first system has taken a definite value (indicated by a strong correlation). 

\subsection{RQM and interactions}\label{interaction}  In an attempt to go beyond the minimal interpretation, one typically faces the question under which conditions a physical quantity may be said to possess a definite value.  For instance, if one knows that a hydrogen atom is in its energy ground state,  one may wish  to say that the energy has then the  value $-13.6$ eV,  the smallest energy eigenvalue of the atom in the appropriate units. More generally, in any state $\rho$, if $p^\sfe_\rho(X)=1$, then one may consistently assume (i.e.\ use the way of speaking without contradictions) that $\sfe$ has a value in $X$,\footnote{In the case of continuous observables, like position and momentum, further care is needed in such a formulation.} 
or, alternatively, 
that the (sharp or unsharp) property $\sfe(X)$ pertains to the system.\footnote{
 Consistently with that, one may also say that a property (sharp or unsharp) $E\in\hE(\hH)$ is objective in a state $\rho$ if either $\tr{\rho E}=1$ or $\tr{\rho (I-E)}=1$. We emphasize that this way of speaking does not mean a commitment to the so-called  `eigenvalue-eigenstate link',  the assumption 
 that if a result (eigenvalue) has been observed, then the state is determined to be the corresponding eigenstate.
} 
To cut short the used vocabulary, we may also refer to an effect $E\ne 0$ as a possible value of an observable $\sfe$ if $E=\sfe(X)$ for some $X\in\hA$, and we recall that any effect $E$ defines a dichotomic observable with the values $E$ and $I-E$, and with the outcome space $\{1,0\}$.

In any state $\rho$ there are observables $\sfe$ and (nontrivial) bin sets $X$ such that $p^\sfe_\rho(X)=1$. 
By the same token,  in any state $\rho$ there also are  plenty of observables $\sff$ with bin sets $Y$ such that $0\ne p^\sff_\rho(Y)\ne 1$.
Facing  this basic situation, RQM poses the old question  and proposes an answer, {\em interaction}: 
\begin{itemize}
\item[(I)]
When and how a probabilistic prediction about the value of a variable [$\sfe$]  of a physical system $S$ is
resolved into an actual value  [$X$]? The answer is this: when $S$ interacts with another physical system $S'$. \cite[p.\ 5]{Rovelli2018}. 
\end{itemize}

Accordingly,
assume  that $S$ interacts with another system  $S'$, with a Hilbert space $\hH'$, 
and assume that the interaction can be described by a unitary operator $U:\hH\otimes\hH'\to\hH\otimes\hH'$.
Given that $S$ and $S'$ are initially, before the interaction, dynamically and probabilistically independent of each other, 
with the respective (pure) states  $\rho=\kb{\fii}{\fii}$ and $\sigma=\kb{\phi}{\phi}$,
the interaction then transforms the initial (unentangled) state 
$\rho\otimes\sigma=\kb{\fii\otimes\phi}{\fii\otimes\phi} $  to the  final, typically entangled state $U(\rho\otimes\sigma) U^*=P[U(\fii\otimes\phi)]$.\footnote{RQM  is somewhat ambiguous with respect to the notion of state, and, in particular, that of an isolated sytem. 
The assumption that   $S$ and $S'$ are initially characterized by the states $\rho$ and $\sigma$ is in accordance with the examples discussed in  \cite{Rovelli1996,Smerlak_Rovelli2007, Laudisa_Rovelli2019, Rovelli2018} and writing 
$\rho\otimes\sigma$ just points  to the fact  that the two systems are  independent of each other (and of the rest of the world).
However, we are also told that  "there is no meaning to ‘the state of an 
isolated system’"  \cite[sec.\ 3.1]{Laudisa_Rovelli2019}.
On the other hand,
Dorato \cite[p.\ 254]{Dorato2016} argues:
"In RQM {\em relata} (isolated quantum systems, or parts) with state dependent dispositional properties ought to be regarded as existent, since there  is no relation without {\em relata}".
} 
 Since the two systems  $S$ and $S'$ are in  a fully symmetric position,\footnote{It is a basic hypothesis of RQM that all systems (including observers) are  treated on the same level, Hypothesis 1 of \cite{Rovelli1996},
a point underlined also by Dorato \cite{Dorato2016} and Pienaar \cite{Pienaar2021}. On the other hand, van Fraassen \cite{BvanF2010},  seems to deviate from this reading of RQM.}
the question arises:
which of the plurality of observables $\sfe$ and $\sfe'$ of $S$ and $S'$, with nointrivial probabilities 
 $0\ne p^\sfe_\rho(X) \ne 1$ and $0\ne p^{\sfe'}_\sigma(Y) \ne 1$, should or could take  values through $U$ with respect to $S'$ and $S$, respectively? 
For instance, $S$ could be a proton and  $S'$ an electron and  the interaction might be the Coulomb one (depending on the mutual distance), the pair $(S,S')$ forming a hydrogen atom.  
Without further qualifications it is difficult to decide which proton variables take values, and which values, with respect to  the electron, and vice versa.
The proposed answer is: 
\begin{itemize}
\item[(SC)]
they are among those observables whose {\em values} get {\em strongly correlated} in the interaction.\footnote{In the primary RQM literature the word {\em correlation} appears without the prefix {\em strong}, see, e.g. \cite[p. 9]{Rovelli1996}. However, in all the discussed examples the  relevant correlations are strong.} 
\end{itemize}
This leads us to consider such correlations.

\subsection{RQM and correlations}\label{correlation}

A preliminary mathematical problem of RQM is thus to determine under which conditions any two effects $E\in\hE(\hH)$ and $F\in\hE(\hH')$ are strongly correlated in the state $U(\rho\otimes\sigma)U^*$. To answer this question, we identify the effects $E$ and $F$ with the dichotomic observables $\{I-E,E\}$ and $\{I-F,F\}$,
with the outcome spaces $\{0,1\}$, and say that they are
strongly correlated in the state  $U(\rho\otimes\sigma)U^*$, if the corresponding
dichotomic probability measures $\mu$ and $\nu$, with $\mu(\{1\})=\tr{\rho^f E}$, and
 $\nu(\{1\})=\tr{\sigma^f F}$, are strongly correlated with respect to their coupling $\gamma$, with $\gamma(\{1,1\})=\tr{U(\rho\otimes\sigma)U^*E\otimes F}$,
where  $\rho^f={\rm tr}_{\hH'}[U(\rho\otimes\sigma)U^*]$ and  $\sigma^f={\rm tr}_{\hH}[U(\rho\otimes\sigma)U^*]$ are the subsystem states, the partial traces of $U(\rho\otimes\sigma)U^*$ over $\hH'$ and $\hH$, respectively.

\begin{remark}\rm To recall the meaning of the above notion, 
consider any two probability measures $\mu$ and $\nu$ (on the real line) and let $\gamma$ be any one of their couplings (joint measures), that is, $\mu(X)=\gamma(X\times\R)$ and $\nu(Y)=\gamma(\R\times Y)$ for all $X,Y$. If one of them is a point measure (Dirac measure), then $\gamma$ is necessarily the product measure, $\gamma=\mu\times\nu$, and their covariance  is 0.
 Assume, thus that neither of them is a point measure. By definition, their normalized correlation coefficient with respect to $\gamma$ is strong (or rather, perfect),  that is, ${\rm cor}(\mu,\nu | \gamma)=\pm 1$, exactly when $\int xy\,d\gamma(x,y)-\int x\,d\mu(x)\int y\,d\nu(y)=\pm \Delta(\mu)\Delta(\nu)$, where $\Delta(\mu),\Delta(\nu)$ are the standard deviations of $\mu,\nu$, and
this is the case exactly when $\mu$ and $\nu$ are completely dependent with a linear  function
$h(y)=ay+b$, that is, $\gamma(X\times Y)=\nu(h^{-1}(X)\cap Y)$, for all $X,Y$,  where $a=\pm\Delta(\mu)/\Delta(\nu)$ and $b=\epsilon_1-a\epsilon_2$, $\epsilon_1=\int x\,d\mu,\epsilon_2=\int y\,d\nu$ ($a>0$ for positive and $a<0$ for negative correlation).
\qed\end{remark}
Any two  effects $E\in\hE(\hH)$ and $F\in\hE(\hH')$ are thus strongly correlated in state $U(\rho\otimes\sigma)U^*$
 exactly when all the probabilities
$\tr{\rho^fE}$, $\tr{\sigma^fF}$, and $\tr{U(\rho\otimes\sigma)U^*E\otimes F}$, are equal.  
Clearly, if the two measures  are point measures then the (sharp or unsharp) properties  are objective in the respective states of $S$, $S'$.

For  two observables $\sfe$  and $\sfe'$ of $S$ and $S'$, the strong correlation of any of their effects $\sfe(X)$ and $\sfe'(Y)$  after the interaction (in state $U(\rho\otimes\sigma)U^*$) is hereby characterized.  
To extend this notion 
to cover all  possible values of the observables, 
it is natural, and, in fact,   necessary, to discretize the observables.
For that end, let $(X_i)$ and $(Y_j)$ be any partitions of the (real) value spaces of $\sfe$ and $\sfe'$ into disjoint bin sets, the so-called reading scales. The values of $\sfe$ and $\sfe'$ with respect to these reading scales 
are then strongly correlated if for each $\sfe(X_i)$ there is an $\sfe'(Y_j)$ such that the pair of effects is strongly correlated in state $U(\rho\otimes\sigma)U^*$.  Finally, discretized or not, the observables $\sfe$ and $\sfe'$ are (by definition) strongly correlated in state $U(\rho\otimes\sigma)U^*$ if the (real) probality measures $p^\sfe_{\rho^f}$ and $p^{\sfe'}_{\sigma^f}$ are strongly correlated with respect to their joint probability measure $\gamma$, with $\gamma(X\times Y)=\tr{U(\rho\otimes\sigma)U^*\,\sfe(X)\otimes\sfe'(Y)}$.

With the assumptions $\rho=\kb\fii\fii$ and $\sigma=\kb\phi\phi$, one may
use a Schmidt decomposition of the state vector $U(\fii\otimes\phi)$
to construct examples of  pairs of effects  and observables whose values are strongly correlated in this state.
Indeed, let
\begin{equation}\label{Schmidt1}
U(\fii\otimes\phi)=\sum_i\lambda_i\sum_{m=1}^{n_i}\xi_{im}\otimes\eta_{im},
\end{equation}
be such  a  decomposition, with $\lambda_i>0,$ $\lambda_i\ne\lambda_j,$ $i\ne j,$ $\sum n_i\lambda_i^2=1$, 
$n_i$ the degeneracy of $\lambda_i$, and
$\ip{\xi_{im}}{\xi_{jn}}=\delta_{ij}\delta_{mn}= \ip{\eta_{im}}{\eta_{jn}}$.
Let
\begin{equation}\label{Schmidt2}
P_i=\sum_m\kb{\xi_{im}}{\xi_{im}}, \qquad
R_i=\sum_m\kb{\eta_{im}}{\eta_{im}},
\end{equation}
so that
\begin{equation}\label{Schmidt3}
\rho^f=\sum_i\lambda_i^2P_i, \qquad \sigma^f=\sum_i\lambda_i^2R_i
\end{equation}
are the spectral decompositions of $\rho^f$ and $\sigma^f$. We note that this decomposition is (essentially) unique\footnote{See, e.g.\ \cite[Theorem 3.15]{QM}.} only when all $n_i=1$. Also, the projections $P_i$, resp.\  $R_i$, need not sum up to $I$, that is, the vectors 
$(\xi_{ij})_{i,j}$, resp.\ $(\eta_{ij})_{i,j}$ need not form an orthonormal basis of $\hH$, resp.\ $\hH'$.

As an immediate observation, 
any (sharp) observables having the  collections of projections $\{P_1,P_2,\ldots\}$ and $\{R_1,R_2\ldots\}$ as  part of their spectral structure have their corresponding `values' $P_i$ and $R_i$ 
strongly correlated in the vector state $U(\fii\otimes\phi)$.  If, for some $k$,  $n_k>1$, that is, the projections $P_k$ and $R_k$ are $n_k$-dimensional, then also any finer observables with those $P[\xi_{km}]$ and $P[\eta_{km}]$ as spectral projections 
are  strongly correlated, too.
Moreover, for any unitary operators $U_k,U'_k$ acting in the subspaces generated by the vectors $(\xi_{kj})_{j}$ and $(\eta_{kj})_{j}$ define other refined observables whose values are now strongly correlated. It is worth noting that the $S$-observables defined by different rotations $U_k$ are mutually incompatible. 
A well-known example of such a case  is given by a pair of qubits brought through an interaction to one of the Bell states.

Though incomplete, the above discussion already shows that strong value correlation is not enough to specify which of the possibly uncountably many observables of $S$ take values with respect to $S'$, and vice versa.
Further information is  needed, and in search for that
we follow an idea expressed  in   \cite[sec.\ 1.3]{Laudisa_Rovelli2019}:
\begin{itemize}
\item[(P)]
In the relational interpretation, any interaction counts as a measurement, to the extent one system affects the other and this influence depends on a variable of the first system. Every physical object can be taken as defining a perspective, to which all values of physical quantities can be referred.
\end{itemize}
The primary RQM literature does not give an explicit definition of  the notion of a perspective but in the examples  discussed they appear as what is called measurement schemes.
This invites us to take a step  to formulate the idea of a {\em perspective} as a {\em measurement scheme}.

\subsection{RQM - perspectives as measurement schemes}\label{scheme}
Let us assume that $S'$  observes $S$, using the interaction $U$ and
fixing  one of its quantities $\sfz$ (with a value space $(\Omega,\hA)$)  as the means of observing or witnessing $S$, that is, as  a pointer or read out observable. 
If $S'$ is able to witness $S$ in each of its initial states $\rho$,\footnote{Less is enough, e.g., pure states suffice.} 
then it is a simple  mathematical fact that the probabilities
$\tr{U(\rho\otimes\sigma)U^*\, I\otimes \sfz(X)},$ $X \in\hA,$ $\sigma$  a fixed state of $S'$, define a unique observable $\sfm$, 
with the   value space $(\Omega,\hA)$,
of $S$ such that these probabilities are those of $\sfm$ in  state $\rho$.\footnote{We take this to  mean that the "influence depends on a variable of the first system". }

In the language of the quantum theory of measurement, system $S'$, with its Hilbert space $\hH'$, the pointer observable $Z$, the initial (`ready') state $\sigma$, and the interaction $U:\hH\otimes\hH'\to\hH\otimes\hH'$, 
in short, $\hM=(\hH',U,\sigma,\sfz)$,
is called a measurement scheme for measuring the observable $\sfm$ on $S$.\footnote{The term measurement scheme, also called premeasurement \cite{BeCL1990,QTM,PM1998},  is  used to emphasize that a full measuring process may require further properties, like a capacity of recording a result. We return to that question in sec.\  \ref{RQMlc}.  We also stress  that the notion of a measurement scheme does not imply any restrictions to the nature of the system $S'$.}
The above mentioned probability reproducability condition can equally be written as $p^\sfm_\rho(X)=p^\sfz_{\sigma^f}(X)$, for all $X,\rho$, where $\sigma^f$ is the final state of $S'$.

Clearly, also $S$ may  witness $S'$, using the same interaction, but fixing one of its observables as a pointer. Again, if $S$ can witness $S'$ in each state of $S'$, then the above can be repeated, but with a changed perspective.
The roles of $S$ and $S'$ are obviously asymmetric in these two processes (as will become even more evident below).

\begin{remark}\rm
It is a fundamental result of quantum mechanics, an application of the dilation theory,  that for each observable $\sfe$ of $S$,
 there is a system $\tilde S$, with a  Hilbert space $\hK$, an interaction $U:\hH\otimes\hK\to\hH\otimes\hK$, a read out observable $\sfz$ (which can be chosen to be sharp) and a ready state $\sigma$ (which can be chosen to be pure) such that the defined observable  is the observable $\sfe$, that is, $p^\sfe_\rho=p^\sfz_{\sigma^f}$ for all $\rho$; 
for a detailed exposition, see, e.g.\ \cite[Chpt 7]{QM}.
Clearly, if we first  fix $S'$, with its Hilbert space $\hH'$, a ready state  $\sigma$, and an  interaction $U$, then  only those observables $\sfe$ of $S$ come into play which, for all $\rho$ and $X$, are  of the form $\tr{\rho\sfe(X)}=\tr{U(\rho\otimes\sigma)U^*\, I\otimes \sfz(X)}$ 
for some $S'$-observable $\sfz$. 
\qed\end{remark}

Let us go on with the assumption that in the interacting pair $(S,S')$, $S'$  is the observing system in  a  pure state $\sigma$, and that
the  interaction  $U$  defines the observable $\sfe$ through a sharp $S'$-observable $\sfz$.
That is, the perspective is defined  by the 4-tuple $\hM=(\hH',U,\sigma,\sfz),$ which defines $\sfe$ in the sense that for any $\rho$, $p^\sfe_\rho(X)=p^\sfz_{\sigma^f}(X)$ for all $X$.
 In general, $\sfe$ is not sharp.\ 
A necessary and sufficient condition for the values $\sfe(X)$ and $\sfz(X)$ to get strongly correlated through $\hM$ is that
all the probabilities $\tr{U(\rho\otimes\sigma)U^*\,\sfe(X)\otimes\sfz(X)}$,   $p^\sfz_{\sigma^f}(X)$,   and $p^\sfe_{\rho^f}(X)$, 
are the same. Since  $p^\sfe_\rho(X)=p^\sfz_{\sigma^f}(X)$, the requirement $p^\sfz_{\sigma^f}(X)=p^\sfe_{\rho^f}(X)$
gives $p^\sfe_{\rho^f}(X)=p^\sfe_\rho(X)$, which is  the first-kind property of $\hM$. Clearly, this is not enough to provide strong value correlations.\footnote{The so-called standard measurement schemes give a plenty of such examples, see, for instance, \cite[sec.\  10.4.]{QM}.}
The additional requirement $\tr{U(\rho\otimes\sigma)U^*\,\sfe(X)\otimes\sfz(X)}=p^\sfz_{\sigma^f}(X)$
is, by $p^\sfz_{\sigma^f}(X)=p^\sfe_\rho(X)$, the repeatability property of $\hM$, which is stronger than the first kind property.
To take up this notion,  we find it useful to recall briefly some relevant aspects of conditional states in measurement schemes.

\begin{remark}\rm 
Consider the state $U(\rho\otimes \sigma)U^*$ and define, for any $X$,  the (unnormalized) state 
\begin{equation}\label{cp1}
I\otimes\sfz(X)U(\rho\otimes\sigma)U^*I\otimes\sfz(X),
\end{equation}
which can be normalized if $\tr{I\otimes\sfz(X)U(\rho\otimes\sigma)U^*I\otimes\sfz(X)}=p^\sfe_\rho(X)\ne 0$. 

Using a characterization of classical conditional probability together with an application of Gleason's theorem, the state \eqref{cp1} can be interpreted as a conditional state, giving rise to conditional probabilities with the condition $I\otimes\sfz(X)$. (For details, see \cite{Cassinelli_Zanghi1983}.) With partial tracing, one gets the corresponding (unnormalized) subsystem states:
\begin{eqnarray}
{\rm tr}_{\hH'}[I\otimes\sfz(X)U(\rho\otimes\sigma)U^*\, I\otimes\sfz(X)] &=&
{\rm tr}_{\hH'}[U(\rho\otimes\sigma)U^*\, I\otimes\sfz(X)] = \hI^\hM(X)(\rho), \label{cp2}\\
{\rm tr}_{\hH}[I\otimes\sfz(X)U(\rho\otimes\sigma)U^*\, I\otimes\sfz(X)] &=&  \sfz(X)\sigma^f\sfz(X).\label{cp3}
\end{eqnarray}
When $p^\sfe_\rho(X)\ne 0$, we let $\rho^f(X)$ and $\sigma^f(X)$ denote the normalized versions of \eqref{cp2} and \eqref{cp3}.

The conditional interpretation of the states $\rho^f(X)$ and $\sigma^f(X)$ is slightly different. Whereas $\sigma^f(X)$ is, in the above Kolmogorov-Gleason  sense, the state of $S'$  after the interaction with the condition $\sfz(X)$, and, clearly,  in that state the property $\sfz(X)$ pertains to $S'$, 
 the interpretation of the state $\rho^f(X)$ is more involved. First of all, the probability $\tr{\rho^f(X)\sfe(X)}$ is not equal to one, in general,  
so that its interpretation as a conditional state of $S$ after the interaction with the condition $\sfe(X)$ is not warranted without further specifications.
We return to that  later. However, for any $S$-observable $\sff$, with a value space $(\Xi,\hB)$,  
the (Kolmogorovian  joint) probability measure $\gamma(Y\times X)={\rm tr}[U(\rho\otimes\sigma)U^*\,\sff(Y)\otimes\sfz(X)]$, $Y\in\hB,$ $ X\in\hA$, is well defined and one has 
\begin{equation}\label{cp4}
p^\sff_{\rho^f(X)}(Y) ={\rm tr}[\rho^f(X)\sff(Y)]=\frac{{\rm tr}[U(\rho\otimes\sigma)U^*\,\sff(Y)\otimes\sfz(X)]}{{\rm tr}[U(\rho\otimes\sigma)U^*\, I\otimes\sfz(X)]}= \frac{\gamma(Y\times X)}{\gamma(\Xi\times X)},
\end{equation}
which shows that  $p^\sff_{\rho^f(X)}(Y)$  is, in the classical sense, the  conditional  probability $\gamma(Y\times\Omega|\Xi\times X)$ of the event $Y\times\Omega $ given the event $\Xi\times X$ with respect to the probability measure $\gamma$. 
In  quantum mechanics with its minimal interpretation, the product 
$p^\sfe_\rho(X)p^\sff_{\rho^f(X)}(Y)=\tr{\hI^\hM(X)(\rho)\sff(Y)}$ is often read as a sequential probability: the probability that a measurement of $\sff$ leads to a result in $Y$ given that a previous measurement of $\sfe$ with $\hM$ in  state $\rho$ led to a result in $X$.\footnote{Here $\hI^{\hM}$ denotes the (completely positive)  instrument, operation valued map, uniquely defined by $\hM$. We wish to underline  that the sequential 
probability is a bi-probability (on pairs of bin sets) and  its interpretation does not presuppose any form of a collapse, that is, it does not presuppose that the state of $S$ would have collapsed, in between,  to the (unnormalized) state $\hI^\hM(X)(\rho)$.}

Classical conditional probability is additive with respect to the disjoint partitions of the conditioning event,  but this is not the case for the Kolmogorov-Gleason notion as applied in equations \eqref{cp1} and \eqref{cp3}. Therefore, if, for instance, $(X_i)$ is a partition of $\Omega$  into disjoint bin sets $X_i$, one has
\begin{equation}\label{cp5}
\rho^f=\hI^\hM(\Omega)(\rho)=\sum_i\hI^\hM(X_i)(\rho)=\sum_ip^\sfe_\rho(X_i)\rho^f(X_i),
\end{equation}
given that $p^\sfe_\rho(X_i)\ne 0$. However, one may also define the Kolmogorov-Gleason conditional state
$\sum_ip^\sfe_\rho(X_i)\sigma^f(X_i)$
 with respect to the partition $(X_i)$  of $\Omega$ (for details, see, \cite{Cassinelli_Zanghi1984}),  but, in general, this conditional state is not $\sigma^f$, that is, the equality
\begin{equation}\label{cp6}
\sigma^f = \sum_ip^\sfe_\rho(X_i)\sigma^f(X_i)
\end{equation}
may fail.

If $\sfz$ is sharp (projection valued) and $\sigma=P[\phi]$ a pure state (as we have  assumed with the choice of the perspective $\hM$), then for any $\rho=P[\fii]$, and for
any  reading scale $(X_i)$, the (validity of the) equality \eqref{cp6} is equivalent to the mutual orthogonality of the component states $\rho^f(X_i)$, that is,  $\rho^f(X_i)\rho^f(X_j)=0$ for all $i\ne j$.
 In this case there is  a strong correlation between the component states $\rho^f(X_i)$ and $\sigma^f(X_i)$. For details, see \cite[Thms 3.11 and 7.2]{BL1996} or \cite[Thm 22.1]{QM}. Still,  the values of $\sfe$ and $\sfz$ with respect to a reading scale $(X_i)$ need not  be strongly correlated. 
\qed\end{remark}

\subsection{RQM -- correlations via repeatability}\label{repeatability}
As already noticed above, a necessary and  sufficient  condition for $\hM$ to produce strong value correlations between the observables $\sfe$ and $\sfz$ is the equality of the probabilities
 $\tr{U(\rho\otimes\sigma)U^*\,\sfe(X)\otimes\sfz(X)}$, $p^\sfz_{\sigma^f}(X)$, and $p^\sfe_{\rho^f}(X)$, for all $X$ and for any $\rho$. 
It turns out that this is exactly the repeatability property of $\hM$.\footnote{PUZZLE 2 of \cite{BvanF2010} expresses some reservations for the use of  repeatable measurement schemes in Rovelli's approach. Due to the equivalence of this notion with the strong value correlations, the same doupt  might equally well concern perspectives (measurement schemes) producing such correlations.}

A measurement scheme $\hM$  is said to be  repeatable if its immediate repetition does not lead to a new result, meaning that the sequential probabilities $p^\sfe_\rho(X)p^\sfe_{\rho^f(X)}(X)$ should be equal  to $p^\sfe_\rho(X)$.
There are many equivalent formulations of this notion (see, for instance, \cite[Def.\ 10.3., Ex.\ 10.9.10]{QM}) and obviously one is: for any $\rho,X$, if $\tr{\rho\sfe(X)}\ne 0$,  then  $p^\sfe_{\rho^f(X)}(X)=1$, that is,
$\tr{U(\rho\otimes\sigma)U^*\,\sfe(X)\otimes\sfz(X)}=\tr{\hI^{\hM}(X)(\rho)\sfe(X)}=p^\sfe_\rho(X)=p^\sfz_{\sigma^f}(X)$. Clearly, this means that
 the property  $\sfe(X)$ pertains to the system in state $\rho^f(X)$,
though it does not mean that the conditional state $\rho^f(X)$  would be a conditional state in the sense of  \eqref{cp3}. For that still further specifications would be  needed, see the below Remark \ref{Davies}.

The repeatability condition is a strong condition implying, in particular,  that the observable  $\sfe$ is  discrete \cite{Ozawa,Luczak1986} (see, also, \cite[Thm 10.4.]{QM}), that is, there is a countable set $\Omega_0=\{\omega_i\,|\, i\in\mathbb I\subset\mathbb N\}\subset \Omega$ such that  each $\{\omega_i\}\in\hA$  and $\sfe(\Omega_0)=I$. 
In that case there is a natural reading scale with $X_i=\{\omega_i\}$. Also, the repeatability implies that both the observables $\sfe$ and $\sfz$ as well as their 
 `values' $\sfe(X_i)$ and $\sfz(X_i)$ are strongly correlated in $U(\rho\otimes\sigma)U^*$ for all $\rho$ \cite[Theorems 5.5.\ and 6.3.]{BL1996}.
If, in addition, the states $\rho^f(X_i)$ are mutually orthogonal, 
then one also has that $\sigma^f=\sum_ip^\sfe_\rho(X_i)\sigma^f(X_i)$, 
meaning  that in this case both of the final states $\rho^f$ and $\sigma^f$ have, in view of the measurement scheme $\hM$,   natural decompositions \eqref{cp5} and \eqref{cp6}  in terms of the final component states $\rho^f(X_i)$ and $\sigma^f(X_i)$  with the relevant weights $p^\sfe_\rho(X_i)$.
This happens, in particular, if 
$\sfe$ is sharp, that is, projection valued.

\begin{remark}\label{Davies}\rm
If in addition to the repeatability, the measurement scheme $\hM$ is also nondegenerate, that is, the possible final states $\{\rho^f\,|\,\rho\ {\rm a\ state}\}$ separate the set of effects, then $\sfe$ is sharp (projection valued).  Further, if  the measurement scheme is also d-ideal, d for discrete, that is, for any $\rho$ and $\omega_i$, if $p^\sfe_\rho(\{\omega_i\})=1$ implies $\rho^f(\{\omega_i\})=\rho$, then $\hM$ is equivalent to a L\"uders measurement of $\sfe$, that is,
$$
\hI^\hM(X)(\rho)=\sum_{\omega_i\in X} \sfe(\{\omega_i\})\,\rho\,\sfe(\{\omega_i\}),
$$
for all $X$ and $\rho$.\footnote{See
\cite[sec.\  4.3]{Davies1976}, or
 \cite[Thm 10.6.]{QM}.}
Clearly, if $\sfe$ is a sharp discrete observable, then 
there is a perspective, a measurement  scheme $\hM$, with the given properties.
In this case, the (unnormalized) state $\hI^\hM(X)(\rho)$ is (also) a Kolmogorov-Gleason conditional state with respect to the partition $(\sfe(\{\omega_i\}))$ of $\sfe(X)$.  Though extensively used in original sources of RQM, 
this  seems not to be an essential requirement of the interpretation. 
%
%
\qed
\end{remark}

According to  RQM,
the {\em strong  value correlation}, and thus the  {\em repeatability} assumption plays   a crucial role  in a process  where a possible  value of an observable gets resolved into an actual value.
These assumptions lead to  the natural decompositions  \eqref{cp5} and \eqref{cp6}. But, as is well known, 
the decompositions do not carry a description of  ignorance on the actual values of the two observables and hence do not easily justify  value assignements, a justification cap known as the objectification or the measurement problem.
(For an extensive discussion of the subject matter, we refer  to \cite{QTM,PM1998},   or \cite[Chpt 22]{QM}).
A final step is still to be taken. Before that two further comments are due.

\begin{remark}\rm
Fixing a perspective $\hM$ to resolve a possible value of a discretized observable $\sfe$, for which $p^\sfe_\rho(X_i)=p^\sfz_{\sigma^f}(X_i)$ for all $i$, does not avoid the fact that the interaction $U$, even if repeatable, nondegenerate and d-ideal  for $\sfe$ in the sense of Remark \ref{Davies}, may create strong correlations (in some states)  also between the values of some other observables of  $S$ and $S'$, observables which may be incompatible with $\sfe$ and $\sfz$. 
Indeed, if the Schmidt decomposition of $U(\fii\otimes\phi)$ contains degenerate eigenvalues $\lambda_i$, then any pointers $\sfz$ 
associated with mutually incompatible rank-1  refinements of the corresponding projections $R_i$  would give such examples.
\qed\end{remark}

Though important,  discrete observables do not exhaust all the observables 
and many realistic measurement schemes, with discrete pointers, define (discrete) unsharp observables.
Neither do repeatable measurements exhaust the measurements of discrete observables.

\begin{remark}\label{R3}\rm
In \cite[sec.\ 2(a)]{Rovelli2018} we find an argument for the discreteness of quantum mechanics:
\begin{itemize}
\item[]
{\em [T]he} major physical characterization of quantum theory is that the volume of the region  $R$
where the system happens to be cannot be smaller that $2\pi\hbar$:
$$
{\rm Vol}(R) \geq 2\pi\hbar,
$$
per each degree of freedom. This is the most general and most important physical fact at the core
of quantum theory. This implies that the number of possible values that any variable distinguishing points within
the region $R$ of phase space, and which can be determined without altering the fact that the system
is in the region $R$ itself, is at most
$$
N \leq \frac{{\rm Vol}(R)}{2\pi\hbar}
$$
which is a {\em finite} number. That is, this variable can take discrete values only. 
\end{itemize}
There is no compact region $R$ of the phase space where the `system  is' or `happens to be' (in the sense that the `values' of the canonical variables would be localized in such an $R$, see the below Example \ref{E2}); thus the first inequality is  trivial $\infty\geq 2\pi\hbar$ and hence  the case for   the finiteness of $N$ does not arise. 
Neither does the so-called unsharp (phase space) localization help in this respect, see, for instance, \cite{PB1984,KLPY2019}.
\qed
\end{remark}

\subsection{RQM and local collapse}\label{RQMlc}
 With the tools collected above we now  try to express the full RQM answer to the question "When and how a probabilistic prediction about the value of a variable  of a physical system is resolved into an actual value?", the proposed answer being "when the system interacts with another system".

Let the totality of the probabilistic predictions concerning the system   $S$ be expressed by a state $\rho$. Let $S'$ be any other system able to define a perspective through a measurement scheme $\hM=(\hH',U,\sfz,\sigma)$ with respect to which values of the quantity $\sfe$ of $S$ can be referred to
in the sense that the predictions  $p^\sfe_\rho(X)$ concerning $\sfe$  are now reproduced as the probabilisties $p^\sfz_{\sigma^f}(X)$ concerning $\sfz$.

 In order to approach the `RQM value assignement'  through strong value correlations, one needs to 
assume that the perspective $\hM$ is repeatable, that is, with $\hM$ we 
also fix a reading scale $(X_i)$, 
with respect to which  the scheme is repeatable (w.r.t.\ $(X_i)$).\footnote{Equivalently, one may choose the pointere observable $\sfz$ to be discrete. Including reading a scale gives more flexibility in considering the possible value assignements of discretized versions of continuous observables like position or momentum.}
In addition, one may assume
that the states $\rho^f(X_i)$ are pairwise orthogonal\footnote{If $\sfe$ is sharp, the orthogonality requirement is redundant.} so that  both $\rho^f=\sum_ip^\sfe_\rho(X_i)\rho^f(X_i)$ and $\sigma^f=\sum_ip^\sfe_\rho(X_i)\sigma^f(X_i)$
as well  as $\tr{\rho^f(X_i)\sfe(X_i)}=1=\tr{\sigma^f(X_i)\sfz(X_i)}$ for all $i$. 
In this case, all the relevant correlations 
are strong. Nevertheless there is no way  to justify that
 the (discretized) observables
$X_i\mapsto\sfe(X_i)$ and  $X_i\mapsto\sfz(X_i)$  would have one of the   possible values  $\sfe(X_i)$ and $\sfz(X_i)$, unless one of the weights $p^\sfe_\rho(X_k)=1$;
all that can be said  on these values  is their strong probabilistic coupling in  the state $U(\rho\otimes\sigma)U^*$,
$$
\tr{U(\rho\otimes\sigma)U^*\,\sfe(X_i)\otimes\sfz(X_j)}=\delta_{ij}p^\sfz_{\sigma^f}(X_j)=\delta_{ij}\,p^\sfe_\rho(X_i).
$$
To close up this  justification  gap RQM poses the following postulate (see, e.g. \cite[p.7]{Rovelli2018}):
\begin{itemize}\label{LC}
\item[(LC)]
   In the $S-S'$\,-interaction   one of the possible ($p^\sfe_\rho(X_i)\ne 0$) values $\sfe(X_i)$ is resolved  into an actual value $\sfe(X_k)$, say, relative to $S'$, and this accompanied with the assumption  that  the state of $S$  with respect to $S'$ is then $\rho^f(X_k)$. 
\end{itemize}
We call this  hypothesis  a {\em local collapse}, a thing which happens to $S$ only with respect to $S'$.\footnote{
Dorato \cite{Dorato2016}  calls this process  a "primitive, mutual manifestation of dispositional properties". Dorato also writes [{\em ibid}.\ p.\ 260]
that
"the manifestation in question ought to be regarded as {\em de facto} irreversible; otherwise no stable measurements would be available".
}

 To underline the {\em ad hoc} nature of this postulate, 
let us recall that $S$ and $S'$ are just any two quantum systems brought into an interaction which correlates strongly the values of the chosen observables.  In particular, there is no detector indicating the occurence of a special result after the interaction.\footnote{See also  Brukner's paper
   "Qubits are not observers" \cite{Brukner2021}.} The postulate (LC) simply stipulates that some $\sfe$-value   occurs in the interaction and due to the strong correlation the state of $S$ w.r.t.\ $S'$ is then an eigenstate of the corresponding $\sfe$-value.\footnote{
This assumption resembles another {\em ad hoc} assumption, called
the eigenvalue-eigenstate link: if one knows the value of a quantity, then  the state has to be a corresponding eigenstate.} 
Though not explicitly mentioned in the original RQM literature, one might expect, in accordance with the below assumption (CPL), Sec.\ \ref{CPL},  that a similar thing happens to $S'$: the pointer $\sfz$ takes the corresponding value, and the state of $S'$ w.r.t.\ $S$ is a corresponding eigenstate.

\subsection{RQM -- sequential perspectives}\label{RQMsp}
From the point of view of the world around $S$, the probabilities for the possible values of  the observables of $S$  (after the interaction)  now vary depending on the perspective: in view of $S'$ they are those coded in the state $\rho^f(X_k)$ whereas for any other system $S''$ they are those coded in $\rho^f$,  that is,  those arising from  $U(\rho\otimes\sigma)U^*$,  which is the state of $S-S'$ w.r.t.\ $S''$.
As emphasized also in  \cite{Rovelli1996,Rovelli2018,Adlam_Rovelli2022}, this means that the observable  $\sfe$ of $S$ could take  different values with respect to other perspectives.

Consider, in sequence, the interactions $S-S_1$ and  $S-S_2$, 
modelled by the schemes $\hM_1,$ and  $\hM_2$, 
respectively, with the initial probabilistic information on $S$ being coded in state $\rho$. 
Since the unitary operator $U_i$ acts nontrivially only in the Hilbert space $\hH\otimes\hH_i$, $i=1,2,$ 
the possibilities for the values of the observables $\sfe_i$ of $S$ defined by the perspectives $\hM_i$ are, in succession, those coded in the states $\rho,$ and $\hI^{\hM_1}(\Omega)(\rho)\equiv\rho^f_1$.\footnote{For simplicity, we omit the possible time evolution between the interventions and assume that the observables $\sfe_i$ have the same outcome space $\Omega$ (usually $\Omega=\R$).}

Let us assume that  the measurement schemes have been tuned to be repeatable with respect to the corresponding reading scales $(X_i)$, and $(Y_j)$.
According to local collapse, in each case some of the possibilities $p^{\sfe_1}_\rho(X_i)\ne 0$, and $p^{\sfe_2}_{\rho^f_1}(Y_j)\ne 0$,
will resolve into actual values, say $\sfe_1(X_l),$ and $\sfe_2(Y_m)$, 
 with the state  of $S$ with respect to $S_1$, and  $S_2$, 
being the normalized versions of $\hI^{\hM_1}(X_l)(\rho)$, and $\hI^{\hM_2}(Y_m)(\hI^{\hM_1}(\Omega)(\rho))$, 
respectively. We may have chosen all the perspectives such that they not only define the same observable, that is, $\sfe=\sfe_1=\sfe_2$, 
 but even so that their instruments are the same, $\hI=\hI^{\hM_1}=\hI^{\hM_2}$.
Due to the repeatability property, the respective local (unnormalized) states of $S$ are then  $\hI^{\hM}(X_l)(\rho)$, 
and  $\hI^{\hM}(Y_m)(\rho)$, 
with the corresponding probabilities $p^\sfe_\rho(X_i)$ for the bin sets $X_l,$ $Y_m$. 
Clearly, the values taken by the observable in question  in this sequence of interventions, need not be the same. 

This is to be compared with the fact that one may also consider the above sequence as a sequential (joint) intervention, with  the following sequential probabilities:
$$
\tr{\hI^{\hM_2}(Y)(\hI^{\hM_1}(X)(\rho))},
$$
with $X,$ $Y,$ 
being the bin sets in the respective value spaces. 
These probabilities define a  biobservable  $\sfe_{12}$  
of $S$ whose possibilities in state $\rho$ are just these probabilities. 
Using the dual instruments, this observable is
\begin{equation*}
\sfe_{12}(X,Y)=\hI^{\hM^1}(X)^*(\hI^{\hM^2}(Y)^*(I))\\
=\hI^{\hM^1}(X)^*(\sfe_2(Y)),
\end{equation*}
with the 
marginal observables
\begin{eqnarray*}
\sfe_{12}(X,\Omega)&=&\sfe_1(X)\\
\sfe_{12}(\Omega,Y)&=&\hI^{\hM^1}(\Omega)^*(\sfe_2(Y)).
\end{eqnarray*}

If, again, the measurement schemes as well as the reading scales are  chosen as above, then 
these probabilities are simply
$$
\tr{\hI^{\hM_2}(Y_j)(\hI^{\hM_1}(X_i)(\rho))}=\tr{\hI^\hM(Y_j\cap X_i)(\rho)}=\delta_{ij}\tr{\hI^\hM(X_i)(\rho)}, 
$$
with $\hI^\hM 
= \hI^{\hM_2}=\hI^{\hM_1}$, and
 $\sfe_{12}(X_i,Y_j)=\sfe(X_i\cap Y_j)=\delta_{ij}\sfe(X_i)$.
Again, by local collapse,  one of these possibilities gets resolved in the process, the
biobservable $\sfe_{12}$
 takes a  value
 $\sfe_{12}(X_k,X_k)$, say, and the state of $S$ with respect to the system $S_1-S_2$ is (the normalized form of)  $\hI^\hM(X_k)(\rho)$.

There  is, however,  no conflict between the two views, since their perspectives are different. In the sequential case, we have a (joint) perspective monitoring the possibilities of the  biobservable $\sfe_{12}$ of $S$  in the state $\rho$, whereas in the first case we have two different perspectives monitoring (separately in succession) the $S$-observables $\sfe_1,$ $\sfe_2,$ in the respective states $\rho,$ $\rho^f_1$.

Finally, we return briefly to the question of correlations coded in the state $U(\rho\otimes\sigma)U^*$ 
arising from a repeatable measurement scheme $(\hH',U,\sigma,\sfz)$ of $\sfe$ (w.r.t.\ a fixed reading scale). Since now
$p^{\sfe\otimes\sfz}_{U(\rho\otimes\sigma)U^*}(X_i,X_j)=0$, for all $i\ne j$, 
any suitable perspective would see (according to RQM) one of the possible pairs $(\sfe(X_i),\sfz(X_i))$  as the `value' of the $S-S'$ observable $\sfe\otimes\sfz$. Clearly, there is no reason to expect that the pair of  values 
taken by $\sfe\otimes\sfz$ with respect to the chosen perspective is the same than the value $\sfe(X_k)$, say,  of $\sfe$ taken by the perspective $(\hH',U,\sigma,\sfz)$, a value strongly correlating with the value $\sfz(X_k)$ of $\sfz$.
However, in \cite[sec.\ 2.6]{Laudisa_Rovelli2019} one reads:
\begin{itemize}
\item[(S)]
Prima facie, RQM may seem to imply a form of perspective solipsism, as the values of variables realized in the perspective of some system $S'$ are not necessarily the same as those realized with respect to another system $S''$. This is however not the case, as follows directly from quantum theory itself. The key is to observe that any physical comparison is itself a quantum interaction. Suppose the variable $\sfe$  of $S$ is measured by $S'$ and stored into the variable $\sfz$ of $S'$. This means that the interaction has created a correlation between $\sfe$ and $\sfz$. 
In turn, this means that a third system measuring $\sfe$ and $\sfz$ will certainly find consistent values. That is: the perspectives of $S'$ and $S''$ agree on this regard, and this can be checked in a physical interaction.
\end{itemize}
In our reading, this is a new independent  assumption and it appears to be incompatible with the preceding ideas trying to exhibit the assumption (LC)  of local collapse.
Pienaar  \cite{Pienaar2021} calls this  the assumption of  {\em shared facts} and he also argues that it is inconsistent with the rest of the RQM assumptions, as understood by him. The claim that  according to RQM "we all ‘see the same world'"\,  \cite[sec.\ 2.6]{Laudisa_Rovelli2019} is not supported by our reading of the underlying assumptions constituting RQM. This conclusion is also in line  with  van Fraassen \cite[sec.\ 5.4]{BvanF2010} who, in his attempt to understand RQM,  
finds it necessary to 
propose an additional postulate in RQM to reach concordance between different observers.

\subsection{RQM -- cross-perspective links}\label{CPL}
To remedy the above defect Adlam and Rovelli have recently completed RQM with 
the postulate of {\em cross-perspective links}
\cite[Definition 4.1]{Adlam_Rovelli2022}, which reads:
\begin{itemize}
\item[(CPL)] In a scenario where some observer Alice [$S'$] measures
a variable $V$  [$\sfe$] of a system $S$, then provided that Alice does not undergo any interactions which
destroy the information about $V$ [$\sfe$] stored in Alice’s physical variables [persumably $\sfz$], if Bob subsequently measures the physical variable [$\sfz$]  representing Alice’s information about the variable $V$ [$\sfe$], then Bob’s [$S''$]
measurement [presumably, with pointer $\sfz'$] result will match Alice’s measurement result.\footnote{The wording of this postulate is somewhat unfortunate in view of the basic ideology of RQM since it may  mislead one to think that the quantum systems $S'$ and $S''$, called Alice and Bob, would possess some extraordinary  properties beyond their pure quantum nature. See also \cite{Brukner2021}.}
\end{itemize}

To spell out this assumption, let $(\hH',U,\sigma,\sfz)$ describe the measurement scheme, called Alice,  of the variable $\sfe$ of $S$ with the probabilities for the possible values of $\sfe$ being coded in state $\rho$ so that $p^\sfe_\rho(X_i)=p^{\sfz}_{\sigma^f}(X_i)$ for any $X_i\in\hR$ (a fixed reading scale), and assume that the value $X_k$, say, has occured. By the local collapse, the state of $S$ with respect to $S'$ is then $\rho^f(X_k)$. On the other hand,
the states of $S$ and  $S'$  with respect to any other system  $S''$ 
should be  $\rho^f$ and $\sigma^f$,  that is, those arising from $U(\rho\otimes\sigma)U^*$ as partial traces. 
Let us assume that Bob,  given as the scheme $(\hH'',U',\pi,\sfz')$,  
now measures the $S'$-variable $\sfz$  using an interaction $U'$ that couples only the systems $S'$ and $S''$.  Thus the probabilities for the possible values of $\sfz$ are coded in the state $\sigma^f$ so that $p^{\sfz}_{\sigma^f}(X_i)=p^{\sfz'}_{\pi^f}(X_i)$ for all $X_i\in\hR$ (Bob using the `same' reading scale).
Note that this is not a sequential measurement of $S$ but  a measurement on $S$ (by $(\hH',U,\sigma,\sfz)$) followed by a measurement on $S'$ (by $(\hH'',U',\pi,\sfz')$).
Since we now have
$$
p^\sfe_\rho(X_i)=p^{\sfz}_{\sigma^f}(X_i)=p^{\sfz'}_{\pi^f}(X_i),
$$
for any $X_i$, the conlusion that "if Alice recorded a result $X_k$ , then the result of Bob has to match with that"\, seems to pressupose that in addition to
the local collapse $\rho\mapsto\rho^f(X_k)$,
the state of $S'$ with respect to $S''$ has to be  collapsed to $\sigma^f(X_k)$.\footnote{Clearly, it would be enough that the state of $S'$ w.r.t.\ $S''$ would be an eigenstate of $\sfz(X_k)$.} It is  a basic feature of  quantum mechanics, there is no way for Bob to get  with certainty the result  Alice recorded unless
the state of $S+S'$ w.r.t.\ $S''$ is an eigenstate of $I\otimes\sfz(X_k)$, that is, 
 the state of $S'$  w.r.t.\ $S''$ is an eigenstate of $\sfz(X_k)$.  The meaning of the provision in  (CPL) remains unclear to us.


This is to be compared with the three different cases discussed in sec.\ \ref{RQMsp}, with assuming that Bob instead performs one of these measurements.
Without repeating the whole discussion, we just note that if
 Bob realises, say,  a joint measurement of $\sfe\otimes\sfz$ on $S-S'$ in state $U(\rho\otimes\sigma)U^*$, any of the possible   results $(X_i,X_i)$ could be obtained, anyone with probability $p^\sfe_\rho(X_i)$.
Without assuming that the postulate of cross-perspective links holds also in this case, the conclusion that "we all `see the same world'" is still unjustified.
In order the result of Bob would match with the result of Alice 
 it is necessary the the state of   $S-S'$, with respect to Bob, $S''$, needs to be such that the result  $(\sfe(X_k),\sfz(X_k))$ is certain, that is,
%
the state of $S-S'$ with respect to $S''$ is such that its partial states are $\rho^f(X_k)$ and $\sigma^f(X_k)$. If one of them is pure, then that state is just the product $\rho^f(X_k)\otimes\sigma^f(X_k)$. Since $S''$ is is arbitrary, this makes the local collapse to be global.


\section{Axiomatic reconstruction and RQM: some remarks}

In his initial 1996 RQM paper, Rovelli bases his approach on two main ideas:
\begin{itemize}
\item[1)]
That the unease [with the text book quantum mechanics] may derive from the use of a concept which is inappropriate to describe the physical world at the quantum level. I shall argue that
this concept is the concept of observer-independent state of a system, or, equivalently, the concept of observer-independent values of physical quantities.
\cite[p.\ 1639]{Rovelli1996}
\item[2)]
That quantum mechanics will cease to look puzzling only when we will be able to derive the formalism of the theory from a set of simple physical assertions (“postulates”, “principles”) about the world. Therefore, we should not try to append a reasonable interpretation to the quantum mechanics formalism, but rather to derive the formalism from a set of experimentally motivated postulates. \cite[p.\ 1639]{Rovelli1996}
\end{itemize}

Since the pioneering works of von Neumann \cite{vN1932} and Birkhoff and von Neumann \cite{BvN1936} a lot of effort has gone 
to searching for a systematic, conceptually clear and mathematically rigorous reconstruction of quantum mechanics from physically motivated axioms. This has led,
 in particular, to the extensive theories of 
 quantum logics  and convex  operational or generalized probabilistic theories. 
From the massive  literature  we mention only 
  the classic monographs of Mackey \cite{Mackey},  Varadarajan \cite{VSV}, Piron \cite{Piron1976}, Mittelstaedt \cite{PM1978}, Beltrametti and Cassinelli \cite{BC},  and Ludwig \cite{Ludwig1983,Ludwig1985}.  
Each of  these approaches,  when appropriately supplemented, leads to a conceptually clear, puzzle-free interpretation of quantum mechanics in accordance with the second of the above two ideas.\footnote{As an example of some such complements, see, for instance \cite{CL2016}.}
The plurality of the existing conceptual frameworks may, however,  suggest that the epic
question "how could the world possibly be the way [quantum] theory says it is?"   \cite[p.\ 4]{BvanF1991}
will not receive a concerted answer.
The powerful framework of generalized probabilistic theories has shown the usefulness of the dual concepts of states (as equivalent classses of preparations) and observables (as equivalent classes of measurements), which  gives weight to the boldness of the first of the above ideas.\footnote{Note, however, that some of the axiomatic approaches, notable those of \cite{Piron1976} and \cite{PM1978}, do not take the notion of a state as a primitive concept. }

In recent years, many of the building blocks of  these approaches have been slightly modified,  reformulated, 
and improved 
to fit better for the needs of  quantum information theory.  Several  reconstructions  of the Hilbert space theory of  finite level systems have thus been achieved, see, for instance,  \cite{Hardy,Mauro,Masanes2011,Masanes,Hohn}. Finite level systems, typically qubits,
 suffice for most of the needs of quantum information technologies but leave out most of quantum physics. It 
 remains to be seen if  such reconstructions could be extended to reconstruct full quantum mechanics,
and thus suport a systematic interpretation of the whole quantum mechanics.

Here we shall briefly comment on some of such reconstructions which derive, at least partly, their motivations from the ideas of RQM.


Instead of developing systematic RQM, Rovelli  outlines the starting point of such a reconstruction along the quantum logic frame 
indicating  two of its basic postulates. In such a frame, a reconstruction is based on a set $L$ of events (called also  questions, or propositions, or experimental functions,  or decision effects, or properties, depending on the point of view) concerning a physical system $S$, and one aims to provide a physically motivated mathematical structure for  $L$ such that, in the end, $L$ could be identified with the projection lattice of an appropriate Hilbert space. 

Without entering into full details, it suffices to recall that  the     two postulates  of  \cite{Rovelli1996} concerning  the "event structure"   are:
\begin{itemize}
\item[] {\bf Postulate 1} (Limited information). There is
a maximum amount of relevant information
that can be extracted from a system;
\item[] {\bf Postulate 2} (Unlimited information). It is
always possible to acquire new information about a system.
\end{itemize}

Assuming that $L$ has the structure of an orthocomplemented lattice, Grinbaum \cite{Grinbaum2005} formalizes  the notion  of 
"relevant information"\, as follows: 
\begin{itemize}
\item[]
an event $b\in L$ is {\em irrelevant} with respect to an event $a\in L$ if $b\land a^\perp\ne 0$; otherwise, $b$ is relevant with respect to $a$, that is, if $b\land a^\perp=0$. 
\end{itemize}
For any $a,b\in L$, if $b\leq a$, then $b$ is relavant to $a$.
We recall that an orthocomplemented lattice is Boolean if  all its  disjoint elements, $a\land b=0$, are also orthogonal,  $a\leq b^\perp$, in which case $a$ is relevant to $b^\perp$ and $b$ is relevant to $a^\perp$. 
In a non-Boolean lattice disjointness does not imply orthogonality and 
one may find mutually incompatible  $a,b\in L $ such that both $b$ and $b^\perp$ are relevant to $a$.

Using  a characterization  of the  orthomodularity 
given in \cite{Maciej}, it is  then shown \cite{Grinbaum2005} that Postulate 1, when interpreted strictly (see below),   implies that $L$ is orthomodular, that is, for any $a,b\in L$, if $a\leq b$ then $b=a\lor(b\land a^\perp)$.  The proof depends  on interpreting ``maximal amount of relevant information'' to mean  that  $L$ is of finite length, that is any maximal chain  $0<a< b<\ldots< 1$ has finitely many elements. The next example shows that this is an essential part of the proof.

\begin{example}\label{E2}\cite[Example 3]{CL2016} \ \rm
 Consider the infinite dimensional vector space  $V=\ell^2(\Q)$ of the square
summable sequences of rational numbers $q = (q_1, q_2, q_3, \ldots)$ with the Hermitian
form $f(q, p) = \sum_{i=1}^\infty q_i p_i$. The lattice $L_f (V)$ of  $f $-closed subspaces is a complete, orthocomplemented,  irreducible AC lattice of infinite length. Since the space $V$ is not complete,  it follows from a theorem of Sol\'er \cite{Soler} together with \cite[Theorem 2.8.]{Piziak1991}, that neither the lattice  $L_f (V)$ nor the space $V$  is  orthomodular.\footnote{A Hermitian space $V$ is orthomodular if for each $M\in L_f(V)$, $M+M^\perp=V$.} 
The vectors
$e_1=(1, 0,\ldots, 0,\ldots),  \ldots,  e_n= (0\ldots, 0, 1, 0, \ldots),\ldots$  
form an orthonormal basis in $V$  and one may construct  an increasing sequence of $f$-closed subspaces 
$V_1=[e_1], V_2=[e_1,e_2], \ldots, V_n=[e_1,e_2,\ldots,e_n],\ldots$
 such that the preceding is relevant with respect to the following one. Here $V_n$ is the  ($n$-dimensional, and thus  $f$-closed) subspace generated by the vectors $e_1,\ldots,e_n$.
Clearly, this sequence is unlimited, showing  that the  above proof of orthomodularity is valid only under the assumption  $L$ has a finite length.

Along with $L_f(\ell^2(\Q))$, the projection lattice $\hP(\hH)$ of a complex separable infinite dimensional Hilbert space $\hH$ contains   plenty of  sequences of mutually relevant events. 
In addition to such ordered sequences, consider, as an example, 
the spectral projections of the (canonical) position and momentum $\sfq$ and $\sfp$. They have the following properties: for any bounded sets $X,$ $Y\in\br$,
$$
\sfq(X)\land\sfp(Y)=\sfq(X)\land\sfp(Y)^\perp=\sfq(X)^\perp\land \sfp(Y)=0,
$$
showing that all such  $\sfp(Y)$  and $\sfp(Y)^\perp$ are relevant to any $\sfq(X)$, and the other way round. 
\qed
\end{example}

This example  suggests that the idea of  "maximal amount of relevant information"\,  of Postulate 1 should, perhaps, be interpreted as "maximal amount of  mutually compatible relevant information"\,  and "maximality"\, 
not to be restricted to the finite case.

As concerns Postulate 2, it  is seen to  imply, together with  Postulate 1,
that $L$ is  non-Boolean, perhaps, even irreducibly so, see also \cite{Trassinelli2018}. 

Despite the attractive nature of the two postulates, the (possible) implication that $L$ is of finite length is ruling out most of quantum physics. Also, though
the orthomodularity and  the non-Boolean character of  $L$ are necessary, there are many alternative ways to get these properties, and, more importantly, 
 there is  still  a long way to reach the Hilbert space realization for the event structure,  see, for instance   \cite{CL2016}. 
Finally,  there  is no  hint how such a reconstruction  might help to justify a relational interpretation of quantum mechanics, especially its core idea of `local collapse', let alone the assumption of `cross-perspective links´.

A systematic information theoretic reconstruction of the  $N$-qubit quantum theory, which derives some of its motivation from relational interpretation,  
is worked out in  great detail in \cite{Hohn,Hohn2}. Instead of following the GPT-approach, the authors construct  a  slightly different general frame of `landscape of gbits theories' $(\mathcal Q,\Sigma,\mathcal T)$, where $\mathcal Q,$ $\Sigma,$ $\mathcal T$ are, respectively,  the sets of all possible (relevant) questions, states, and time evolutions describing the physical system $S$ under consideration. The authors pose 4 \cite{Hohn2} plus 1 \cite{Hohn} principles of information acquisition  which determine the structure of the triple $(\mathcal Q,\Sigma,\mathcal T)$ to be the familiar Hilbert space theory based on the  $N$-fold tensor product of the qubit Hilbert space $\C^2$.
The  first two of these principles are conceptually motivated by the above  two postulates of Rovelli, whereas the other three formulate the requirements of information preservation (in-between interrogations), time evolution (of the catalog of knowledge, states), and question unrestrictedness (the physical realizability of the (relevant) questions). Even though this approach shares many features of RQM, there is, however, a fundamental difference: in RQM all  the interacting systems $S$ and $S'$ are treated on the same basis as quantum systems, whereas here the observing system $S'$ is a classical system (not described), able to detect the answers to the posed questions.  A possible value of a physical quantity of the system $S$ is here resolved into an actual value through a classical observer's detection, an idea contradicting the very starting point of RQM.
The so-called measurement problem is hereby `resolved' by simply assuming that the state of the system $S$ after interrogating a question `collapses' according to the yes/no --answer to the relevant `eigenstate'.

Finally, an axiomatic reconstruction of quantum mechanics gives us mathematically coherent and  conceptually clear, puzzle-free interpretations of quantum mechanics. Still, there remains the question of how the abstract Hilbert space theory should be applied in concrete cases, like atoms and molecules, or, say, photons.  To the best of our knowledge, the most succesful approach starts with first characterising the Poincar\'e and the Galilei invariant isolated and elementary quantum objects within the abstract Hilbert space theory
and then extends this analysis to mutually interacting objects, 
a method initiated by the seminal papers of Wigner \cite{Wigner1939}, Bargmann \cite{Bargmann1954}, and Mackey \cite{Mackey1958}, and further developed, for instance in the  monographs of Mackey \cite{Mackey1989}, Varadarajan \cite{VSV}, and Ludwig \cite{Ludwig1983}, as well as in Cassinelli {\em et al}.\ \cite{C2004}.  Such an approach does not easily fit with the basic ideology of RQM.

\

\centerline{*********************}

In {\em Helgoland:  making sense of the quantum revolution} we  read that "[r]elational QM is a consistent interpretation of quantum theory". 
In  our reading  of the basic assumptions of the relational interpretation of quantum mechanics, which goes along 
 the  studies of  van Fraassen \cite{BvanF2010}, Laudisa \cite{Laudisa2019}, and Pienaar \cite{Pienaar2021}, 
such a conclusion appears premature. With adding  the postulate of cross-perspective links, Adlam and Rovelli \cite{Adlam_Rovelli2022} conclude that  the thus updated RQM
"guarantees intersubjective agreement between observers when they perform measurements on
one another". 
This may well be the case. However, in our reading, the price is a somewhat implicit return to the controversial projection postulate which is a crude way to avoid the measurement or objectification problem of quantum mechanics. 
We may thus close with the words of Laudisa \cite{Laudisa2019} that
 "we have reasons to be much less optimistic toward the prospects of RQM: 
a lot of work needs to be done before RQM may aspire to become a satisfactory interpretational 
framework for the main foundational issues in QM".

\

\noindent
{\bf Acknowledgements.} We thank  both  Carlo Rovelli and Jacques Pienaar for their separate comments on an earlier version, the version arXiv:2207.01380, of our paper. The comments of JP led to  some reformulations  in Sections \ref{RQMlc} and \ref{CPL}  but we were unable to match the comments of CR. 
We thank also Jukka Kiukas for valuable discussions in the course of this work.


\begin{thebibliography}{99}
\bibitem{Rovelli1996} C. Rovelli, Relational Quantum Mechanics, 
{\em  Int. J. Theor. Phys.} {\bf 35} (1996) 1637. arXiv:quant-ph/9609002.
\bibitem{Smerlak_Rovelli2007} M. Smerlak, C. Rovelli, Relational EPR, 
{\em Found. Phys.} {\bf 37} (2007).

\bibitem{Laudisa_Rovelli2019} F. Laudisa, C. Rovelli, Relational Quantum Mechanics, {\em Stanford Encyclopedia of Philosophy},
First published Mon Feb 4, 2002; substantive revision Tue Oct 8, 2019.

\bibitem{Rovelli2018} C. Rovelli, Space is blue and birds fly
through it,  {\em Phil. Trans. R. Soc.} {\bf  A 376} (2017).

\bibitem{Martin_Rovelli_etal_2019} P. Martin-Dussaud, C. Rovelli,  F. Zalamea,
The Notion of Locality in Relational Quantum Mechanics,
{\em Found. Phys.}  {\bf 49} (2019) 96–106.


\bibitem{Rovelli2021}C. Rovelli, {\em Helgoland: making sense of the quantum revolution}, Riverhead Books, Penguin Random House LLC, 2021. Original, in Italy, 2020.

\bibitem{Adlam_Rovelli2022} E. Adlam, C. Rovelli, Information is Physical: Cross-Perspective Links in Relational
Quantum Mechanics, arXiv:2203.13342v2.

\bibitem{BvanF2010} Bas C. van Fraassen, 
Rovelli's World, 
{\em Found. Phys.} {\bf 40} (2010) 390–417. 

\bibitem{Ruyant2018} Q. Ruyant,
Can We Make Sense of Relational Quantum Mechanics?, 
{\em Found. Phys.} {\bf 48} (2018) 440–455.




\bibitem{Laudisa2019} F. Laudisa, Open Problems in Relational Quantum Mechanics, 
 {\em J. Gen. Philos. Sci.}  {\bf 50} (2019) 215–230 . 

\bibitem{Pienaar2021} J. Pienaar, A Quintet of Quandaries: Five No‑Go Theorems 
for Relational Quantum Mechanics,  {\em Found. Phys.} {\bf 51} (2021) 97. 
https://doi.org/10.1007/s10701-021-00500-6.

\bibitem{Brukner2021} \v C. Brukner, Qubits are not observers – a no-go theorem, 
arXiv:2107.03513 (2021).


\bibitem{Rovelli_DiBiagio2021} A. Di Biagio, C. Rovelli,  Relational Quantum Mechanics is About Facts, Not States: A Reply to Pienaar and Brukner. Found Phys 52, 62 (2022). https://doi.org/10.1007/s10701-022-00579-5
 arXiv:2110.03610v1. 


\bibitem{QM} P. Busch, P. Lahti, J.-P. Pellonp\"a\"a, K. Ylinen, {\em Quantum Measurement}, Springer, 2016.

\bibitem{CDeVL1997} G. Cassinelli, E. De Vito, A. Levrero, On the decompositions of a quantum state, {\em J. Math. Anal. Appl,} {\bf 210} (1997) 472-483.

\bibitem{Dorato2016} M Dorato, Rovelli's relational quantum mechanics, anti-monism, and quantum becoming, in {\em The Metaphysics of Relations}, eds. A. Marmodoro, D. Yates, Oxford UP, 2016, pp.\ 235-261.

\bibitem{BeCL1990} E. Beltrametti, G, Cassinelli, P. Lahti, Unitary measurements of discrete quantities in quantum mechanice, {\em J. Math. Phys.} {\bf 31} (1990) 91-96.

\bibitem{QTM} P. Busch, P. Lahti, P. Mittelstaedt, {\em The Quantum Theory of Measurement}, {\bf LNP m2}, Springer, 1991, 2nd rev. ed. 1996.

\bibitem{PM1998} P. Mittelstaedt, {The Interpretation of Quantum Mechanics and the Measurement Process}, Cambridge UP, 1998.


\bibitem{Cassinelli_Zanghi1983} G. Cassinelli, N. Zanghi,  Conditional probabilities in quantum mechanics. I. Conditioning with respect to a single event. {\em  Nuovo Cim. B} {\bf 73} (1983) 237-245.

\bibitem{Cassinelli_Zanghi1984} G. Cassinelli, N. Zanghi,  Conditional probabilities in quantum mechanics. II.  Additive conditional probabilities. {\em  Nuovo Cim. B} {\bf 79} (1984) 141-154.


\bibitem{BL1996} P. Busch, P. Lahti, Correlation properties of quantum measurements, {\em  J. Math. Phys.} {\bf 37}  (1996) 2585-2601.

\bibitem{Ozawa} M. Ozawa, Quantum measuring processes of continuous observables, {\em J. Math. Phys.} {\bf 25}  (1984) 79-87.

\bibitem{Luczak1986} A. \L uczak, {\em Instruments on von Neumann algebras}, Institute of Mathematics, \L \'od\'z University, 1986.

\bibitem{Davies1976} E.B. Davies, {\em Quantum Theory of Open Systems}, Academic Press, 1976.



\bibitem{PB1984} P.  Busch, On joint lower bounds of position and momentum observables in quantum mechanics, {\em  J. Math.
Phys.}  {\bf 25} (1984) 1794–7 .

\bibitem{KLPY2019} J. Kiukas, P. Lahti, J.-P. Pellonp\"a\"a, K. Ylinen, Complementary Observables in Quantum Mechanics, 
{\em Found. Phys.} {\bf 49} (2019) 506.






\bibitem{BCL1990} P. Busch, G. Cassinelli, P. Lahti, On the Quantum Theory of Sequential Measurements, {\em  Found. Phys.} {\bf 20} (1990) 757-775. 

\bibitem{vN1932} J. von Neumann, {\em Mathematische Grudlagen der Quantenmechanik}, Springer 1932, 2. Auflage 1996.

\bibitem{BvN1936} G. Birkhoff, J. von Neumann, The Logic of Quantum Mechanics, {\em Ann. Math.} {\bf 37} 823 (1936).

\bibitem{Mackey} G. Mackey, {\em Mathematical Foundations of Quantum Mechanics}, Benjamin, 1963.

\bibitem{VSV} V.S. Varadarajan, {\em Geometry of Quantum Theory}, Springer, 1985, First published in 1968.

\bibitem{Piron1976} C. Piron, {\em Foundations of Quantum Physics}, Benjamin, 1976.

\bibitem{PM1978} P. Mittelstaedt, {\em Quantum Logic}, D. Reidel, 1978.

\bibitem{BC} E. Beltrametti, G. Cassinelli, {\em The Logic of Quantum Mechanics}, Cambridge UP, 2010, First published in 1981.

\bibitem{Ludwig1983} G. Ludwig, {\em Foundations of  Quantum Mechanics I}, Springer, 1983.

\bibitem{Ludwig1985} G. Ludwig, {\em An Axiomatic Basis for Quantum Mechanics, Derivation of Hilbert Space Structure}, Springer, 1985.


\bibitem{CL2016} G. Cassinelli, P. Lahti, An Axiomatic Basis for Quantum
Mechanics, {\em Found. Phys.} {\bf 46} (2016) 1341-1373.

\bibitem{BvanF1991} Bas C. van Fraassen, {\em Quantum Mechanics: an empiricist view}, Clarendon, 1991. 

\bibitem{Hardy} L. Hardy, Quantum theory from five reasonable axioms,
arXiv:quant-ph/0101012. Reconstructing quantum theory, arXiv:1303.1538.


\bibitem{Mauro}  G. M. D’Ariano, G. Chiribella, P. Perinotti, 
{\em Quantum Theory from First Principles: an informational approach}, Cambridge UP, 2017.


\bibitem{Masanes2011} L. Masanes, M.P. M\"uller, A derivation of quantum theory from physical
requirements, {\em New Journal of Physics} {\bf  13} (2011) 063001.

\bibitem{Masanes} L. Masanes, M.P. M\"uller, R. Augusiak, D.P\' eres-Garc\' ia, Existence of an information unit as a postulate of
quantum theory, Proceedings of the National Academy of Sciences, {\bf 110} (2013) 16373-16377.

\bibitem{Hohn} P.A. Höhn, C.S.P. Wever, Quantum theory from questions, {\em Phys. Rev. A} {\bf 95} (2017) 012102.

\bibitem{Grinbaum2005} A. Grinbaum, Information-theoretic principle entails orthomodularity of a lattice, {\em Foundations of Physics Letters} {\bf 18} (2005) 563-572.

\bibitem {Maciej} M. J. M\c acynski, On a functional representation of the lattice of
projections on a Hilbert space,  {\em Studia Mathematica} {\bf  47} (1973) 253–259.

\bibitem{Soler} P. Sol\'er, Characterization of Hilbert spaces by orthomodular spaces, {\em Comm. Algebra} {\bf 23} (1995) 219-243.

\bibitem{Piziak1991} R. Piziak,  Orthomodular lattices and quandratic spaces: a survey, {\em Rocky Mt J. Math}. {\bf 21} (1991) 951–992.

\bibitem{Trassinelli2018} M. Trassinelli, Relational Quantum Mechanics and Probability, {\em Found Phys}  {\bf 48 } (2018) 1092–1111. 

\bibitem{Hohn2}  P. A. H\"ohn, Toolbox for reconstructing quantum theory from rules on information acquisit, 
{\em Quantum} {\bf 1}  (2017) 38. 


\bibitem{Wigner1939} E. Wigner,  Unitary representations of the inhomogeneous Lorentz group, {\em Annals of Mathematics} {\bf 40} (1939) 149-204.

\bibitem{Bargmann1954} V. Bargmann, On unitary ray representations of continuous groups, {\em Annals of Mathematics} {\bf 59}  (1954) 1-46.

\bibitem{Mackey1958} G. Mackey, Unitary representations of group extensions. I, {\em Acta Mathematica} {\bf 99}  (1958) 265-311.

\bibitem{Mackey1989} G. Mackey, {\em Unitary Group Representations in Physics, Probability, and Number Theory}, Addison-Wesley, 1989.

\bibitem{C2004} G. Cassinelli {\em et al}, {\em The Theory of Symmetry Actions in Quantum Mechanics}, Springer, LNP 654, 2004.




\end{thebibliography}
\end{document}